\documentclass[prb,twocolumn,showpacs,amsmath,amssymb,superscriptaddress]{revtex4-2}
\usepackage{dcolumn}
\usepackage{bm,graphicx}
\usepackage{etoolbox}
\usepackage{url}
\usepackage{booktabs}
\usepackage[colorlinks]{hyperref}
\usepackage{breakurl}
\usepackage{color}
\usepackage [latin1]{inputenc}

\begin{document}

\title{Temperature dependence of the energy band gap in ZrTe$_5$: \\implications for the topological phase}

\author{I.~Mohelsky}
\affiliation{Laboratoire National des Champs Magn\'etiques Intenses, LNCMI-EMFL, CNRS UPR3228, Univ.~Grenoble Alpes, Univ.~Toulouse, Univ.~Toulouse~3, INSA-T, Grenoble and Toulouse, France}

\author{J.~Wyzula}
\affiliation{Department of Physics, University of Fribourg, Chemin du Mus\'ee 3, 1700 Fribourg, Switzerland}

\author{B.~A.~Piot}
\affiliation{Laboratoire National des Champs Magn\'etiques Intenses, LNCMI-EMFL, CNRS UPR3228, Univ.~Grenoble Alpes, Univ.~Toulouse, Univ.~Toulouse~3, INSA-T, Grenoble and Toulouse, France}

\author{G.~D.~Gu}
\affiliation{Condensed Matter Physics and Materials Science Division, Brookhaven National Laboratory, Upton, New York 11973-5000, USA}

\author{Q.~Li }
\affiliation{Condensed Matter Physics and Materials Science Division, Brookhaven National Laboratory, Upton, New York 11973-5000, USA}
\affiliation{Department of Physics and Astronomy, Stony Brook University, Stony Brook, NY 11794-3800, USA}

\author{A.~Akrap}
\affiliation{Department of Physics, University of Fribourg, Chemin du Mus\'ee 3, 1700 Fribourg, Switzerland}

\author{M.~Orlita}
\email{milan.orlita@lncmi.cnrs.fr}
\affiliation{Laboratoire National des Champs Magn\'etiques Intenses, LNCMI-EMFL, CNRS UPR3228, Univ.~Grenoble Alpes, Univ.~Toulouse, Univ.~Toulouse~3, INSA-T, Grenoble and Toulouse, France}
\affiliation{Faculty of Mathematics and Physics, Charles University, Ke Karlovu 5, Prague, 121 16, Czech Republic}
    
\date{\today}

\begin{abstract}
Using Landau level spectroscopy, we determine the temperature dependence of the energy band gap in zirconium pentatelluride (ZrTe$_5$). We find that the band gap reaches $E_g=(5 \pm 1)$~meV at low temperatures and increases monotonously when the temperature is raised. This implies that ZrTe$_5$ is a weak topological insulator, with non-inverted ordering of electronic bands in the center of the Brillouin zone. Our magneto-transport experiments performed in parallel show that the resistivity anomaly in ZrTe$_5$ is not connected with the temperature dependence of the band gap. 
\end{abstract}

\maketitle

Zirconium pentatelluride (ZrTe$_5$) is a narrow-gap system widely explored in the past~\cite{FurusethACS73,FjellvagSSC86,SmontaraPhysBC86,IzumiJPCCM87},
often in the context of so-called resistivity anomaly~\cite{OkadaJPSJ80, DiSalvoPRB81, SkeltonSSC82,TrittPRB99,WangPRL21,GourgoutnjpQM22}. More recently, theoretical studies of ZrTe$_5$ that were primarily focused on topological properties induced a renewed interest in this system. In particular, the ZrTe$_5$ monolayer was proposed 
to be a large-gap quantum spin Hall insulator~\cite{WengPRX14}. Over the past few years, a number of experiments has been performed on bulk and thin layers of ZrTe$_5$ and expanded our knowledge about this material considerably. The experimentally observed phenomena in ZrTe$_5$ comprise the magneto-chiral effect \cite{LiNP16,WangPRL22}, the giant planar Hall effect \cite{LiPRB18}, the quasi-quantized quantum Hall effect~\cite{TangNature19,GaleskiNC21}, pressure-induced superconductivity~\cite{ZhouPNAS16}, or optical conductivity that is linear in photon energy~\cite{ChenPRB15,MartinoPRL19}. 

This series of important experimental observations contrasts with our current understanding of low-energy excitations in ZrTe$_5$. In fact, no consensus has been so far established whether the electronic bands in bulk ZrTe$_5$ have normal or inverted ordering, \emph{i.e.}, whether this material is a strong or weak topological insulator (STI or WTI). Ab initio calculations favor the STI phase~\cite{FanSR17}, but their prediction force may be limited in a system close to the topological phase transition. Experimentally, both STI and WTI phases have been reported, 
see Refs.~\cite{ManzoniPRL16,ChenPNAS17,MutchSA19,XuPRL18,WangNatureComm21,WangPRL22} and~\cite{MoreschiniPRB16,WuPRX16,Xiang-BingPRL16,XiongPRB17,ZhangNC17}, respectively,  
based on transport, ARPES and optical data.

In this work, we explore, using THz/infrared magneto-spectroscopy, the temperature dependence of the band gap in ZrTe$_5$, in order to test the topological phase of this material. We find that the gap is non-zero at low temperatures, $E_g=(5\pm1)$~meV at $T=5$~K, and it increases with $T$ monotonously. Such behaviour is consistent with ZrTe$_5$ being a weak topological insulator. 

\begin{figure}[t]
      \includegraphics[width=0.45\textwidth]{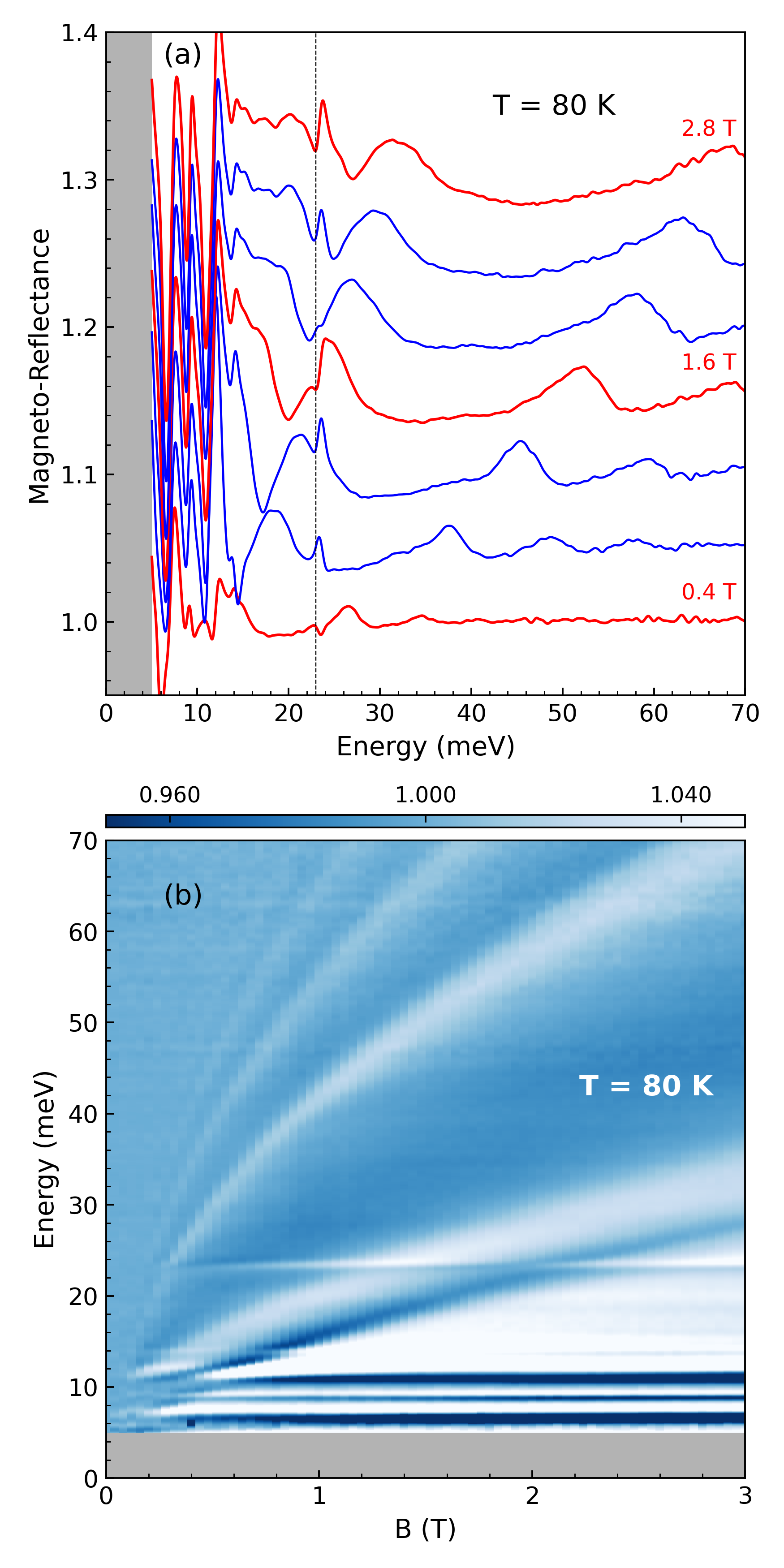}
      \caption{\label{Fig1} (a) Stacked-plot of relative magneto-reflectivity spectra, $R_B(\omega)/R_0(\omega)$, collected on ZrTe$_5$ at $T=80$~K and plotted for selected values of $B$. (b) the false-color plot of $R_B/R_0$ plotted with a step of 40~mT. Weakly $B$-dependent features, nearly horizontal in the plot are phonons, accompanied at low energies (below 10~meV) by an interference pattern. This appears since ZrTe$_5$ becomes transparent at low energies when a sufficiently strong magnetic field is applied. The vertical line in (a) at 23~meV shows the position of the most pronounced phonon mode in ZrTe$_5$~\cite{ChenPRB15}.}
\end{figure}

Our experiments were performed on bulk ZrTe$_5$ synthesized using self-flux growth~\cite{LiNP16}. The explored sample was attached using GE varnish to a metallic holder that was covered by a thin paper sheet, and surrounded by the helium exchange gas.  The temperature was controlled locally, using a heating coil, and measured by two Cernox thermometers placed at two different locations nearby the sample. A relatively large thickness of the sample ($\approx$0.5~mm) was chosen  to avoid effects related to the temperature-induced strain. Results obtained on this sample, presented here, were reproduced on several other specimens with different thicknesses, and glued in different configurations. 

We measured relative magneto-reflectivity, $R_B/R_0$, in the Faraday configuration, \emph{i.e.}, with the wave vector of light propagating parallel to the magnetic field aligned with the $b$ crystallographic axis. The radiation of a globar was analyzed by the Bruker Vertex 80v Fourier-transform spectrometer and delivered via light-pipe optics to the sample located in a superconducting coil. A part of the reflected signal was deflected towards an external bolometer, using a silicon beamsplitter. In addition, magneto-transport data were collected in-situ, \emph{i.e.}, in parallel with magneto-optical measurements. To this end, electric contacts were created using graphite paste in a standard 4-contact configuration.

An illustration of the collected  magneto-reflectivity data is presented in Fig.~\ref{Fig1}. It comprises relative magneto-reflectivity, $R_B/R_0$, measured at $T=80$~K plotted (a) in a form of a stacked-plot of spectra taken at selected values of $B$, and 
(b) as a false-color map. The observed response is in line with preceding 
magneto-optical studies of ZrTe$_5$~\cite{ChenPRL15,ChenPNAS17,JiangPRB17,MartinoPRL19}.  
It closely resembles the magneto-optical response of Landau-quantized electrons in a conical band~\cite{OrlitaSST10}, nevertheless, the characteristic $\sqrt{B}$ dependence of inter-LL excitations is here only approximate. The observed transitions extrapolate to a non-zero energy in the limit of vanishing $B$, thus indicating a band gap in ZrTe$_5$.

\begin{figure*}[h!]
      \includegraphics[width=0.9\textwidth]{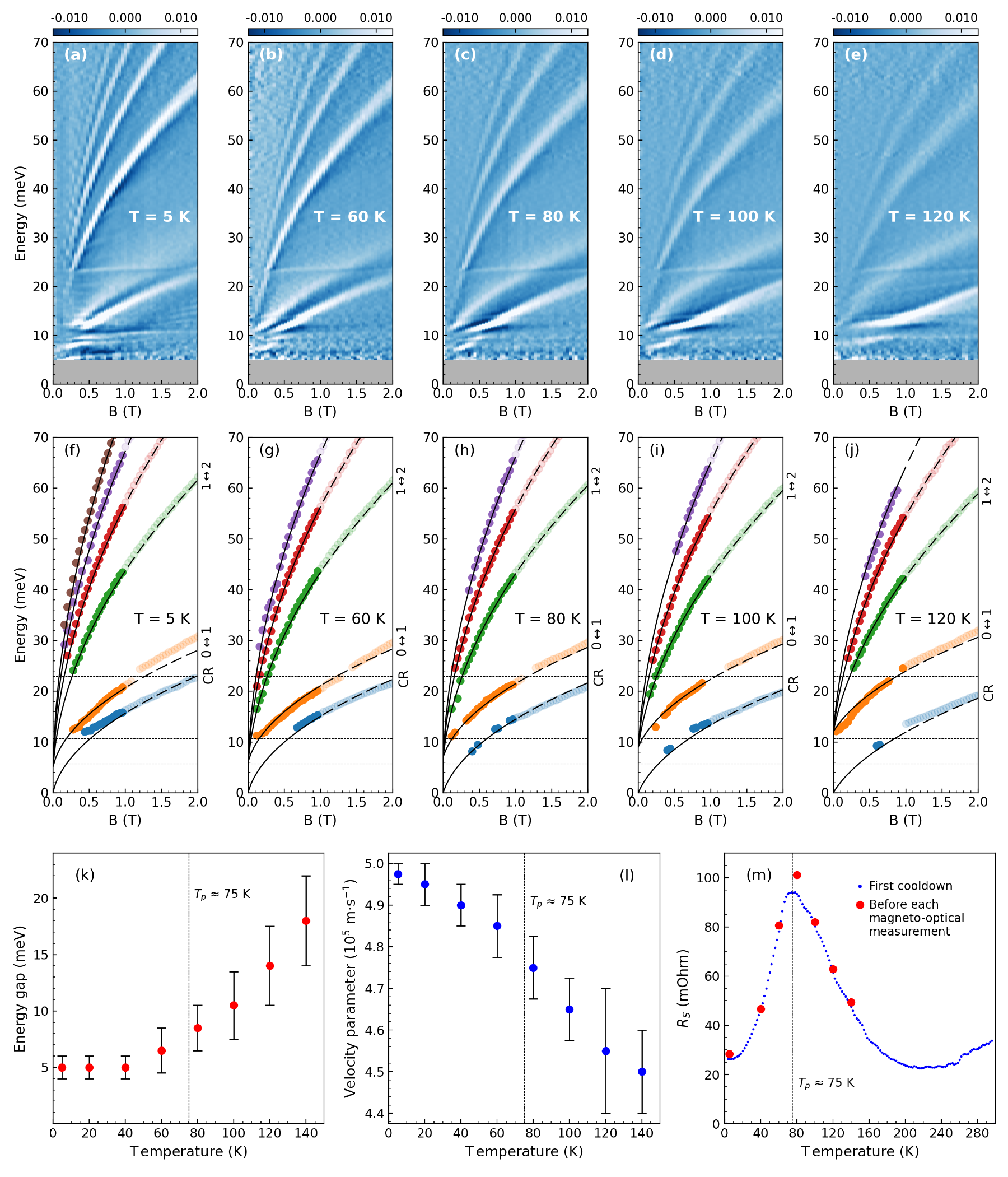}
      \caption{\label{Fig2} (a-e) False-color plots of the first field-derivative of relative magneto-reflectivity, $d/dB [R_B/R_0]$, measured at the temperature of $T=5, 60, 80, 100$ and 120~K. (f-j) The extracted positions of resonances at a given magnetic field and temperature. 
      The solid lines represent the fits of interband inter-LL transitions for the data collected below 1~T. The dashed lines represent extrapolations of these fits to higher magnetic fields and energies. Dotted horizontal lines show energies of three most pronounced phonon modes (at 6, 11 and 23~meV) observed in Refs.~\cite{ChenPRB15,MartinoPRL19}.       (k) The extracted energy band gap of ZrTe$_5$ as a function of temperature. (l) The experimentally determined temperature dependence of the effective velocity parameter. (m) The resistance anomaly in the explored ZrTe$_5$ sample, with the maximum of $R_S(T)$ around $T_p\approx 75$~K. The blue dots show the sample resistance during the first cool-down, the red circles before each run of the magneto-optical measurements at a given $T$. The temperature corresponding to the maximum in resistivity is marked by the vertical dashed lines in (k-m).}
\end{figure*}

To extract the band gap, we have plotted and examined the first field-derivative of the collected spectra, $d/d B[R_B/R_0]$, (Fig.~\ref{Fig2}a-e) and associated the corresponding maxima with energies of individual inter-LL resonances. This approach is applicable for absorption lines on a positive dielectric background. When additional dissipative processes appear in the background, the position of the resonance may be shifted from the inflection point of $R_B(\omega)/R_0(\omega)$ towards the maximum. This uncertainty increases the error bar of the below extracted/discussed values of the velocity parameter, but has a minor influence on the width of the band gap --  the main target of our analysis. 

To understand the observed response in the simplest possible manner, we neglect the 
relatively flat dispersion of electrons along the magnetic field ($b$ axis in our case), see Ref.~\cite{MartinoPRL19} and interpret our data in terms of excitations in the spectrum 
of two-dimensional Landau-quantized massive Dirac electrons: $E_n^\pm=\pm\sqrt{\Delta^2 + v^2 2e\hbar n B}$ where $n=0,1,2\ldots$. Here $v$ stands for the velocity parameter averaged in the plane perpendicular to $\mathbf{B}$ ($a$-$c$ plane) and $+(-)$ refers to the conduction (valence) band. The band gap corresponds to the value of $E_g=2\Delta$. 

The conventional selection rules for electric-dipole transitions, $n\rightarrow n\pm1$, imply the existence of a series of interband inter-LL excitations: $E^+_{n+1(n)}-E^-_{n(n+1)}=\sqrt{\Delta^2+v^2 2e\hbar B n}+\sqrt{\Delta^2+v^2 2e\hbar B(n+1)}$. In our data, such transitions are visible for $n=0,1,2,3$ and 4. In addition, cyclotron resonance (CR) absorption appears. In a weakly electron-doped sample, only the fundamental CR mode of electrons ($0^+\rightarrow1^+$) is active at low temperatures, at the energy of $\sqrt{\Delta^2+v^22e\hbar B}-\Delta$. At elevated temperatures, the fundamental CR mode of holes ($1^-\rightarrow 0^-$) emerges as well, at the same photon energy. 
In contrast to a preceding study~\cite{JiangPRL20}, we did not observed any series of excitations that would indicate another gap. The outlined massive Dirac model was used to fit the low-$B$ part of the data (below 1~T), see Fig.~\ref{Fig2}f-j. This is due to Zeeman effect, which scales linearly with $B$ and which starts to impact visibly the magneto-optical response of ZrTe$_5$ at higher magnetic fields, see Ref.~\cite{ChenPNAS17,JiangPRB17}. 

Despite the apparent simplicity of the massive Dirac model, all observed transitions can be well-reproduced in the chosen range of low magnetic fields and also for all explored 
values of $T$ (see Figs.~\ref{Fig2}f-j). Certain deviations appear in the region when transitions approach phonon lines (horizontal dotted lines in Figs.~\ref{Fig2}f-j) which are excluded from the fitting procedure. This allows us to extract the temperature dependence of $v$ and $\Delta$ which are the only adjustable parameters of the used model (see Figs.~\ref{Fig2}k and l). The extracted values evolve smoothly and monotonously with temperature. The velocity parameter decreases slightly, which likely reflects only the unit cell increasing in size. This in turn implies a decrease
in the overlap of atomic functions, and consequently, overall flattening of the band structure. The band gap gradually grows from the low-temperature value, $2\Delta= (5\pm1)$~meV,  which is consistent with the preceding magneto-optical studies~\cite{ChenPNAS17,MartinoPRL19,JiangPRB17}.

Interestingly, having established that the massive Dirac model is valid, one may infer the width of the band gap $2\Delta$, as well as its increase with $T$, directly from the data in Figs.~\ref{Fig2}a-e without any fitting procedure. It simply corresponds to the separation between the lowest interband line ($0^-\rightarrow1^+$ and $1^-\rightarrow0^+$) and the fundamental CR mode ($0^+\rightarrow1^+$ and $1^-\rightarrow0^-$). Other methods, presumably with a smaller resolution in energy, such as the ARPES or STM/STS techniques, provide us with the band gap larger by a factor of 3 or more~\cite{XiongPRB17,ZhangSR21,Xiang-BingPRL16,WuPRX16}.

With increasing $T$, and above $T=100$~K in particular, the observed interband inter-LL transitions visibly weaken, see Figs.~\ref{Fig2}a-e. This is partly due to thermally induced broadening of LLs, partly, when $k_BT$ exceeds $\Delta$, due to charge carriers excited thermally to LLs with higher indices. The latter reduces the transition strengths via the corresponding occupation factor. At low energies, the optical response becomes also modified by thermally excited carriers. With increasing $T$, the plasma edge forms gradually~\cite{MartinoPRL19} and the simple one-electron picture is then no longer applicable. Instead, the observed CR mode starts to resemble, see Fig.~\ref{Fig2}e, 
the upper CR branch in the classical magneto-plasma response~\cite{PalikRPP70}: $\hbar(\omega_c^2+\omega_p^2)^{1/2}$, where $\omega_p$ and $\omega_c$ are the screened plasma frequency and bare single-electron cyclotron frequency, respectively.

Let us now discuss implications of our experimental findings. Theoretical studies~\cite{WengPRX14,FanSR17,MonserratPRR19} suggest that bulk ZrTe$_5$ is a system close to a topological phase transition, which may, in fact, be the reason why no consensus about the topological nature of ZrTe$_5$ has so far been achieved. Theoretically, only a relatively small increase in the unit-cell volume is sufficient to bring ZrTe$_5$ from the STI to WTI regime, \emph{i.e.}, from the inverted to the normal ordering of bands at the center of the Brillouin zone. Importantly, the band gap always closes and re-opens during such a topological phase transition.

Experimentally, it was shown that the unit-cell volume in ZrTe$_5$ increases monotonously with temperature~\cite{FjellvagSSC86}. Thanks to this, the temperature dependence of the band gap was proposed~\cite{FanSR17,MonserratPRR19} to be an unambiguous signature of the topological phase in ZrTe$_5$. In the STI regime, the band gap is first supposed to shrink with increasing $T$, then close completely, and finally, re-open in the WTI regime. In contrast, a monotonous increase of the band gap with temperature is expected in the WTI regime. Only the latter scenario is consistent with our data (Fig.~\ref{Fig2}g), thus implying the WTI phase in ZrTe$_5$. Let us note that the volume of the unit cell changes considerably in the explored range of temperatures, approximately by 0.5\%~\cite{FjellvagSSC86}. 

Additional magneto-transport measurements, performed along with the magneto-reflectivity experiments, allowed us to correlate the evolution of the band gap width with the well-known
resistivity anomaly in ZrTe$_5$. This effect refers to  a characteristic peak in resistivity appearing when the temperature increases, see, \emph{e.g.}, Refs.~\cite{OkadaJPSJ80, DiSalvoPRB81, SkeltonSSC82,TrittPRB99,WangPRL21,GourgoutnjpQM22}. The temperature dependence of resistance of the explored sample measured during the first cool-down of the sample, $R_S(T)$, is plotted in Fig.~\ref{Fig2}m along with $R_S$ values measured prior to the sweeps of our magneto-reflectivity measurements at a given fixed $T$. We speculate that the small discrepancies between the resistance values in the subsequent experiments, particularly visible at the peak value, could be due to a thermal-cycle-induced change in the contact resistance. This would modify the sourcing current, assumed to be constant while extracting resistance value. The observed profile, reaching maximum around $T_p\approx 75$~K is perfectly in-line with preceding experiments on samples with similar doping. In literature~\cite{XuPRL18,TianPRB19}, the appearance of resistivity anomaly in ZrTe$_5$ was sometimes associated with the temperature-driven closure of the band gap around $T_p$. Our data do not support such a conclusion. 

In summary, we have extracted the temperature dependence of the band gap in ZrTe$_5$ using Landau level spectroscopy. We have found that the band gap increases monotonously, from the 
low-temperature value of $E_g=(5\pm 1)$~meV to nearly 20~meV at $T=140$~K. The observed behaviour is consistent with ZrTe$_5$ being a weak topological insulator.

\begin{acknowledgments}
The work has been supported by the ANR projects DIRAC3D (ANR-17-CE30-0023) and COLECTOR (ANR-19-CE30-0032). A.A. acknowledges funding from the  Swiss National Science Foundation through project PP00P2\_202661. The authors also acknowledge the support of the LNCMI-CNRS in Grenoble, a member of the European Magnetic Field Laboratory (EMFL). The work at BNL was supported by the US Department of Energy, office of Basic Energy Sciences, contract no. DOE-sc0012704. This research was supported by the NCCR MARVEL, a National Centre of Competence in Research, funded by the Swiss National Science Foundation (grant number 205602).
\end{acknowledgments}

\emph{Note added.} During the review process, we learned about another magneto-optical study of ZrTe$_5$ study~\cite{JiangCM22} presenting similar experimental results but suggesting a different interpretation.


%

\end{document}